\documentclass[proceedings, preprint]{rmaa}

\usepackage{paralist}

\usepackage{psfrag,color}

\SetYear{2008}
\SetConfTitle{IV Reuni\'on sobre Astronom\'\i{}a Din\'amica en Latino Am\'erica}

\title{Milli-arcsecond Binaries} 

\author{
  R. M. Torres,\altaffilmark{1} 
  L. Loinard,\altaffilmark{1} 
  A. Mioduszewki,\altaffilmark{2}
  and L. F. Rodr\'\i{}guez\altaffilmark{1}}

\altaffiltext{1}{Centro de Radioastronom\'\i{}a y Astrof\'\i{}sica,
  UNAM, Campus Morelia, Apartado Postal 3--72, 58090, Morelia,
  Michoac\'an, M\'exico (r.torres@astrosmo.unam.mx).}

\altaffiltext{2}{National Radio Astronomy Observatory, Array Operations Center,
  1003 Lopezville Road, Socorro, NM 87801, USA (amiodusz@nrao.edu).}

\shortauthor{Torres et al.}
\shorttitle{RevMexAA(SC) Demo Document}

\listofauthors{R. M. Torres, L. Loinard, A. Mioduszewki, \&  L. F. Rodr\'\i{}guez}
\indexauthor{Torres, R. M.}
\indexauthor{Loinard, L.}
\indexauthor{Mioduszewki, A.}
\indexauthor{Rodr\'\i{}guez, L. F.}

\abstract{As part of an astrometric program, we have used the Very
Long Baseline Array to measure the trigonometric parallax of several
young stars in the Taurus and Ophiuchus star-forming regions with
great accuracy. Additionally, we have obtained an unprecedented sample
of high-resolution ($\sim$ 1 mas) images of several young stellar
systems. These images revealed that about 70\% of the stars in our
sample are very tight binary stars (with separations of a few mas).
Since it is highly unlikely that 70\% of all stars are such tight
binaries, we argue that selection effects are at work.}

\resumen{Como parte de un programa astrom\'etrico, hemos estado usando
el Very Long Baseline Array para medir la paralaje trigonom\'etrica a
varias estrellas j\'ovenes en las regiones de formaci\'on estelar de
Tauro y Ophiuchus con muy alta precisi\'on. Adicionalmente, hemos
obtenido una muestra de im\'agenes de muy alta resoluci\'on ($\sim$ 1
mas) sin precedentes. Estas im\'agenes revelan que alrededor de un
70\% de las estrellas en nuestra muestra son estrellas binarias muy
cercanas (con separaciones de unos cuantos milisegundos de
arco). Debido a que es altamente improbable que 70\% de todas las
estrellas sean binarias con separaciones tan peque\~nas, en \'este
trabajo discutimos los efectos de selecci\'on en el proyecto.}

\addkeyword{astrometry}
\addkeyword{binaries: general}
\addkeyword{radiation mechanisms: nonthermal}
\addkeyword{stars: pre-main sequence}
\addkeyword{stars: individual (V773 Tau AB)}

\begin{document}
% Typeset article header
\maketitle

\section{Introduction}
\label{sec:introduction}

In the last few decades, studies of the solar neighbourhood have led
to the conclusion that most stars show a high multiplicity rate:
$\sim$ 55\% for solar-type stars (Duquennoy \& Mayor 1991), and $\sim$
35 to 42\% for M-dwarfs (Reid \& Gizis 1997; Fischer \& Marcy
1992). Young stars also exhibit a high multiplicity rate: $\sim$ 50\%
with separations of between 0.02 and 1 arcseconds (K\"ohler \& Leinert
1998; Duch\^ene et al.\ 2004; Konopacky et al.\ 2007). Indeed, several
star-forming regions appear to show a multiplicity rate even higher
than that of main sequence stars (Prosser et al.\ 1994; Duch\^ene et
al.\ 1999; Bouvier et al.\ 1997). This shows that the multiplicity
must be already established in the very early phases of
star-formation. It is still a matter of debate, however, if different
environmental conditions of star formation lead to different degrees
of multiplicity.

Observations have shown that non-thermal, high-energy processes often
take place in young stellar objects, particularly in weak-line T Tauri
stars (WTTS). These phenomena produce compact X-ray and radio emission
(Strom \& Strom 1994; Feigelson \& Montmerle 1999). In this project,
we will focus on T Tauri stars known to exhibit such strong
non-thermal activity.

\section{Observations}
\label{sec:observations}

We have chosen from the literature a list of 9 young stellar objects:
5 in the Taurus complex, and 4 in the Ophiuchus complex. Those sources
are low-mass stars (with the exception of S1 which is a main sequence
B star). All 9 sources were previously known to be non-thermal radio
emitters, and had been detected with VLBI techniques in the past (see
Table 1). We will make use of several series of continuum 3.6 cm (8.42
GHz) Very Long Baseline Array observations of each source. The data
were edited and calibrated in standard fashion using the Astronomical
Image Processing System (AIPS).

\section{Results}
\label{sec:results}

\begin{table*}[!t]\centering
  \newcommand{\DS}{\hspace{6\tabcolsep}}
  \setlength{\tabnotewidth}{0.9\textwidth}
  \setlength{\tabcolsep}{1.33\tabcolsep}
  \tablecols{8}
  \caption{Observed stellar systems characteristics in the project}
  \label{tab:observed}  
\begin{tabular}{cccccccc}
\hline
\hline
\\[-0.3cm]
{Region} & {Source} & {Phase\tabnotemark{a}} & {Spectral} & \multicolumn{4}{c}{Members\tabnotemark{b}}\\%
\cmidrule(l){5-8}
{      } & {      } & {Evolution           } & {Type    } & {Known} & {Detected} & {New} & {Total}\\%
\\[-0.3cm]
\hline
\hline
\\[-0.3cm]
{Taurus} & T Tauri Sb  & CTTS      & M0-1    & 3 & 1 & 0 & 3\\%
         & Hubble 4    & WTTS      & K7      & 1 & 1 & 0 & 1\\%
         & HDE 283572  & WTTS      & G5      & 1 & 1 & 0 & 1\\%
         & V773 Tau AB & CTTS/WTTS & K2, K7  & 4 & 2 & 0 & 4\\%
         & HP Tau/G2   & WTTS      & G0      & 3 & 1 & 0 & 3\\%
\\[-0.3cm]
\hline
\\[-0.3cm]
{Ophiuchus} & S1      & MS        & B5 V & 2 & 2 & 0 & 2\\%
            & DoAr 21 & WTTS      & K0   & 2 & 2 & 1 & 3\\%
            & VSSG 14 & WTTS      & A5-7 & 2 & 2 & 1 & 3\\%
            & WL 5    & CTTS/WTTS & F7   & 1 & 3 & 2 & 3\\%
\\[-0.3cm]
\hline
\hline
  \tabnotetext{a}{CTTS: classical T Tauri star, WTTS: weak-lined T
  Tauri star, MS: main sequence star.}  \tabnotetext{b}{From column 5
  to 8 are listed the members known before our observations, the
  members that we detected in non-thermal radio emission, the new
  detected members, and the total members for each system after our
  observations.}
\end{tabular}
\end{table*}

In this work we find that $\sim$ 70\% of the stars in our sample are
multiple stellar systems with separations of a few
milli-arcseconds. In the Ophiuchus complex we found a new member in
both DoAr 21 and VSSG 14, and two new members in WL 5 (Loinard et al.\
2008; see Table 1 and Fig.\ 1). S1 was alredy known to be a binary
(Richichi et al.\ 1994) and we did detect the two members in some of
our images. In the case of Taurus, we knew that T Tauri Sb was a
member of a triple system, but of these three members we only detect
one in non-thermal radio emission (Loinard et al.\ 2005, 2007; see
Table 1). V773 Tau AB was previously known to be a quadruple system,
and we detect two of the four members. Hubble 4 and HDE 283572, as
well as HP Tau/G2 are apparently single (Torres et al.\ 2007,
2008). It is interesting that the fraction of tight binaries appears
to be larger in Ophiuchus than in Taurus.

Note that 1 milli-arcsecond at the distance of Taurus (137 pc; Torres
et al.\ 2007) and Ophiuchus (120 pc; Loinard et al.\ 2008) are $\sim$
0.137 and $\sim$ 0.120 AU, respectively.

In addition to the high multiplicity rate, we found extended emission
in most sources of both complexes (see Fig.\ 1). This might be related
to the extents of the magnetosphere. Young stellar objects often
contain powerful flares and strong magnetic activity, that might
explain the extended emission.

\section{Discussion}
\label{sec:discussion}

\subsection{Binarity}
\label{sec:binarity}

\begin{figure}[!t]
\centering
\setlength{\fboxrule}{0 pt}
\fbox{\includegraphics[angle=0,width=0.8\columnwidth]{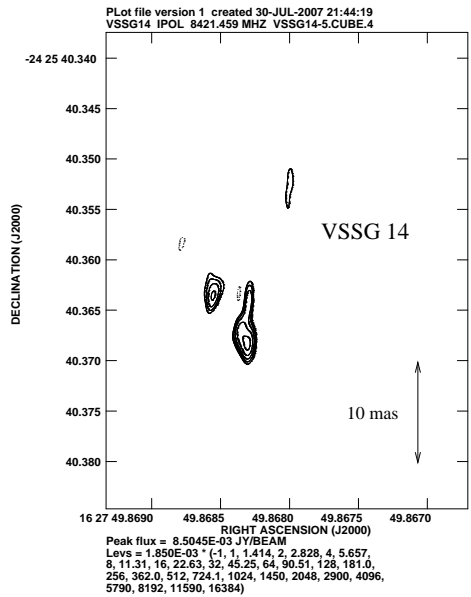}}
\caption{VSSG 14, multiple stellar system in Ophiuchus complex. Note
the extended structure.}
\end{figure}

It is quite unlikely that 70\% of all stellar systems in the Solar
neighborhood are tight binaries such as those found in our VLBA
observations. The high binary rate found here, is instead likely the
result of a selection effect. The systems considered in this project
were selected because they were known to be non-thermal emitters that
had been detected with VLBI techniques before. So the high binary rate
may indicate that tight binaries are more likely to emit non-thermal
radio emission than looser binaries or single stars. This idea is
reinforced by the observations of V773 Tau AB (Fig.\ 2 and 3) where we
find that the emitted flux is a clear function of the separation
between the two stars (see below).

\subsection{V773 Tau}
\label{sec:v773tau}

\begin{figure*}
\centering
\setlength{\fboxrule}{0 pt}
\fbox{\includegraphics[angle=0,width=0.8\textwidth]{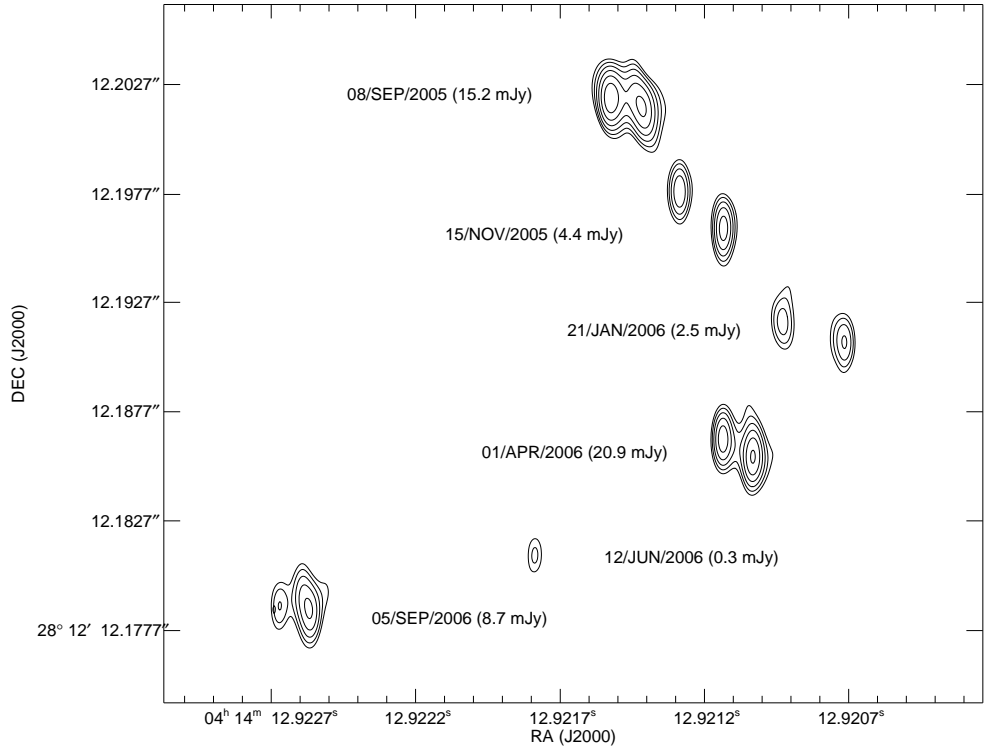}}
\caption{3.6 cm VLBA images of V773 Tau AB at six different
epochs. The separation between V773 Tau A and V773 Tau B changes from
epoch to epoch, and the flux density changes significantly with
separation.}
\end{figure*}

V773 Tau (HD 283447) is the strongest stellar radio source in the 5
GHz VLA survey of WTTS in the Taurus-Auriga molecular cloud complex
(O'Neal et al.\ 1990). It was shown to be a triple system when it was
almost simultaneously found to be a spectroscopic binary with an
orbital period of 51.075 days (Welty 1995), and to have a companion at
about 150 mas (Ghez et al.\ 1993, Leinert et al.\ 1993). A fourth
stellar source was recently found in the system (Duch\^ene et al.\
2003), so V773 Tau appears to be a young quadruple stellar system,
with the spectroscopic binary (V773 Tau A and V773 Tau B) at the
center, and the two companions V773 Tau C and V773 Tau D in wider
orbit.

We have been collecting 19 VLBA observations of the compact
non-thermal radio source associated with the spectroscopic binary V773
Tau A+B to measure its parallax and proper motions. The system almost
always happened to be resolved into two sources. In Fig.\ 2 we show
the binaries at the first 6 epochs. Note that as the separation
between the sources varies from epoch to epoch, the brightness of the
source increases from 0.3 mJy to 21 mJy.

The total flux is shown as a function of time in Fig.\ 4, where the
bars indicate successive periastron passages. The brightness of the
source increased from 0.3 mJy (fifth epoch) to 55 mJy (epoch 16). The
flux density clearly tends to increase when the sources are near
periastron. This is consistent with the results reported by Massi et
al.\ (2002), who also found periodic radio flaring.

While the physical mechanisms leading to this behavior is not entirely
clear, it certainly suggests a relation between radio flux and
separation.

\begin{figure*}
\centering
\setlength{\fboxrule}{0 pt}
\fbox{\includegraphics[angle=0,width=\textwidth]{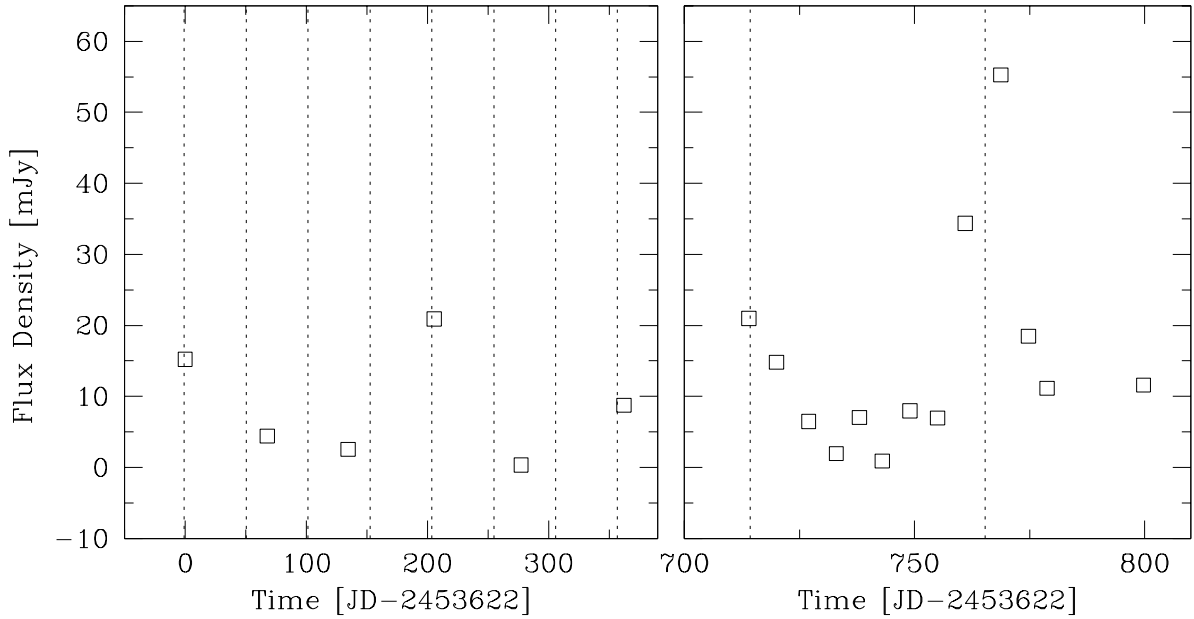}}
\caption{Flux evolution of V773 Tau AB. The flux is shown as boxes,
the sizes of wich represent the magnitudes of errors. The bars in the
figure show the periastron passages assuming as initial epoch
t$_{0}$=2449330.94 JD (Welty 1995)}
\end{figure*}

\section{Conclusions and Perspectives}
\label{sec:conclusions}

We used the Very Long Baseline Array to obtain an unprecedent sample
of high-resolution images of young stellar systems. We found that 70\%
of the stars in the sample are multiple stellar systems with
separations of a few milli-arcseconds. We argue that the high binary
rate indicates that tight binaries are more likely to emit non-thermal
radio emission than looser binaries or single stars; this ideas
reinforced by the observations of V773 Tau AB where we find that the
flux density tends to increase when the sources are near periastron.

The mass is the most fundamental parameter of a star, because it
determines its structure and evolution. For pre-main sequence stars,
there are no reliable empirical mass determinations (mass estimates
are usually based on comparisons with theoretical evolutionary
models). For the binary stars detected here, a direct mass
determination will be possible using Kepler's third law and the
measured relative motions. We are currently obtaining new observations
for those sources that were found to be multiple. These observations
followed the sources during complete orbital periods, and will allow
the determination of the orbital parameters, and therefore of the
masses.

\acknowledgements{R.M.T, L.L., and L.F.R. acknowledge the financial
support of DGAPA, UNAM and CONACyT, M\'exico. NRAO is a facility of
the National Science Foundation operated under cooperative agreement
by Associated Universities, Inc.}

\end{document}